\documentclass[aps,prb,twocolumn,superscriptaddress,english]{revtex4}
\usepackage{amssymb}
\usepackage{graphicx}
\usepackage{amsmath}
\usepackage{soul}
\usepackage{amsthm}
\usepackage{amsfonts}
\usepackage[T1]{fontenc}
\usepackage[latin9]{inputenc}
\usepackage{array}
\usepackage{multirow}
\usepackage{color}
\usepackage{esint}
\usepackage{bm}
\usepackage{color}
\usepackage{bbm}
\usepackage{hyperref}
\usepackage{babel}
\usepackage{titlesec}

\newcommand{\braket}[1]{\left\langle{#1}\right\rangle}
\newcommand{\nh}{{\mathrm{nh}}}

\newcommand{\tot}{{\mathrm{tot}}}
\newcommand{\Tr}{{\mathrm{Tr}}}

\begin{document}

\global\long\def\id{\mathbbm{1}}
\global\long\def\ui{\mathbbm{i}}
\global\long\def\ud{\mathrm{d}}
\title{Manipulating the Relaxation Time of Boundary-Dissipative Systems through Bond Dissipation}
\author{Yi Peng}
\thanks{These authors contribute equally to this work.}
\affiliation{International Quantum Academy, Shenzhen 518048, China}
\author{Chao Yang}
\thanks{These authors contribute equally to this work.}
\affiliation{Shenzhen Institute for Quantum Science and Engineering,
	Southern University of Science and Technology, Shenzhen 518055, China}
\affiliation{International Quantum Academy, Shenzhen 518048, China}
\affiliation{Guangdong Provincial Key Laboratory of Quantum Science and Engineering, Southern University of Science and Technology, Shenzhen 518055, China}
\author{Yucheng Wang}
\thanks{Corresponding author: wangyc3@sustech.edu.cn}
\affiliation{Shenzhen Institute for Quantum Science and Engineering,
	Southern University of Science and Technology, Shenzhen 518055, China}
\affiliation{International Quantum Academy, Shenzhen 518048, China}
\affiliation{Guangdong Provincial Key Laboratory of Quantum Science and Engineering, Southern University of Science and Technology, Shenzhen 518055, China}
\begin{abstract}
	Relaxation time plays a crucial role in describing the relaxation processes of quantum systems. We study the effect of a type of bond dissipation on the relaxation time of boundary dissipative systems and find that it can change the scaling of the relaxation time $T_c\sim L^{z}$ from $z=3$ to a value significantly less than $3$. We further reveal that the reason such bond dissipation can significantly reduce the relaxation time is that it can selectively target specific states. For Anderson localized systems, the scaling behavior of the relaxation time changes from an exponential form to a power-law form as the system size varies. This is because the bond dissipation we consider can not only select specific states but also disrupt the localization properties. Our work reveals that in open systems, one type of dissipation can be used to regulate the effects produced by another type of dissipation.
\end{abstract}
\maketitle

\section{Introduction}
Relaxation processes of quantum systems interacting with their environments are among the most foundational nonequilibrium phenomena. A piece of material in contact with baths at its two boundaries can reach a nonequilibrium steady state, corresponding to the simplest scenario of nonequilibrium systems~\cite{LandiRMP}. In recent years, with the development of experimental techniques providing us with various highly controllable platforms to study the dynamics of open quantum systems~\cite{Bloch,Diehl,Syassen,Weimer,Lee,Tomadin2011,Ludwig,Viciani,Maier,YingHu,Dongning,XueP}, significant advances have also been made in the study of quantum systems coupled to different baths at their edges~\cite{LandiRMP,Prosen2014,Prosen2008,Karevski,Popkov,Guo2021,Goold,Vicari,Carollo,Saha,Clark2019,Clerk2024,PRE2015}.  A pivotal inquiry here pertains to determining the timescale for a boundary-dissipative system to attain a steady state. 
The Liouvillian gap remains an important quantity for characterizing the relaxation time. Except in some special cases~\cite{PRE2015,Mori2021,Bensa2022,Mori2020,Lee2023,ZeqingWang,Ueda2021}, the relaxation timescale can usually be estimated by the inverse of the Liouvillian gap~\cite{PRE2015,Ueda2021,Prosen,ZCai,Bonnes2014,Shibata2019}. Previous results have shown that for various boundary-dissipated systems, the Liouvillian gap $\Delta_g$ scales with the system length $L$ as $\Delta_g\sim L^{-3}$ for integrable systems~\cite{PRE2015,Prosen2008,Yamanaka2023,Schen,VicariPRA,SYZhang} and $\Delta_g\sim e^{-L/l}$ for Anderson localization (AL) systems~\cite{Prosen,Zhou2022exp}, with $l$ being the localization length.
Consequently, the corresponding scaling of the relaxation time is $T_c\sim L^{3}$ and $T_c\sim e^{L/l}$, respectively.

In this work, we investigate how to manipulate the relaxation time of a quantum system with boundary dissipation. This issue holds significant relevance in applications, as regulating the relaxation time is essential for transport properties, quantum control, and information processing. Apart from the particle number decay at the boundaries, we introduce a type of bond dissipation that can be realized experimentally. We find that it can change the scaling of the relaxation time $T_c\sim L^{z}$ from $z=3$ to a value of $z$ significantly less than $3$. In other words, it can significantly reduce the relaxation time, allowing the system to reach equilibrium more quickly. When this type of dissipation is applied to a localized system with boundary dissipation, the scaling behavior of the relaxation time changes from an exponential form to a power-law form as the system size varies. 
We further elucidate the mechanism by which this bond dissipation reduces the relaxation time. Since this type of dissipation can be experimentally realized, it can be used to regulate the relaxation time of boundary-dissipative systems.

\section{Model and results}
We consider the simplest one-dimensional model with only nearest-neighbor hopping, whose Hamiltonian is 
\begin{equation}\label{freeH}
	H_0=-J\sum_{m=1}^{L-1}\left(c_{m+1}^\dagger c_m +c_m^\dagger c_{m+1}\right),
\end{equation}
where $c_m$ ($c_{m}^\dag$) is the annihilation (creation) operator for a particle at site $m$,
and $J$ is the hopping amplitude between neighboring sites, which is set to $1$ as the unit energy.  We consider a boundary-dissipative system where the particle loss operator acts only on the first and last sites of the lattice. This can be described by a purely imaginary on-site potential,
\begin{equation}\label{imagV}
	V_\mathrm{nh}=-i\gamma\left(c_1^\dag c_1+c_L^\dag c_L\right).
\end{equation}

We then introduce bond dissipation acting on a pair of sites $m$ and $m+\ell$, described by
~\cite{Diehl,PZoller2,PZoller3,PZoller4,WangYC,Marcos,Yusipov,Yusipov2,Thaga}
\begin{equation}\label{noise}
	D_m 
	= \left(c_m^\dag+ac_{m+\ell}^\dag\right)
	\left(c_m-ac_{m+\ell}\right),
\end{equation}
where $a=1$ or $-1$, $\ell=1$ or $2$, and $m=1,\ldots,L-\ell$. This type of dissipation can be realized through cold atoms in optical superlattices~\cite{Diehl,PZoller2,PZoller3,PZoller4,WangYC} or through arrays of superconducting microwave resonators~\cite{Marcos,Yusipov}. This operator obviously does not change the particle number, but it does alter the relative phase between this pair of sites separated by a distance $\ell$. 
They are synchronized from an out-of-phase mode to an in-phase mode (or vice versa) by this operator when $a$ is set to $1$ (or $-1$). 

\begin{figure}[b]	
	\centering
	\includegraphics[width=0.485\textwidth]{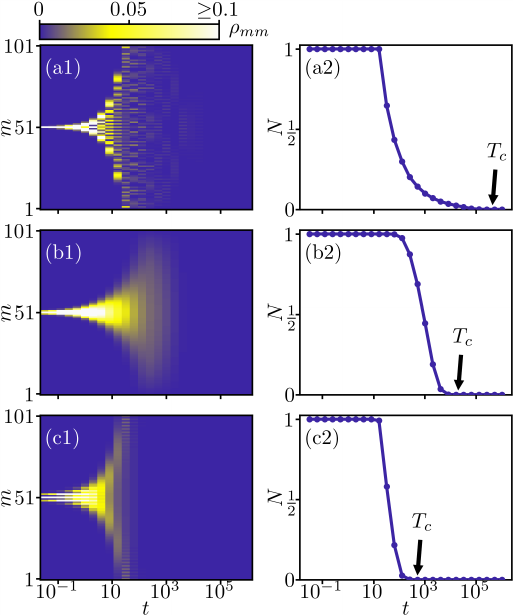}
	\caption{The occupation probability $\rho_{mm}$ of each lattice site [(a1), (b1), (c1)] and the total number of particles $N$ [(a2), (b2), (c2)] change over time for (a1) and (a2) $\Gamma=0$; (b1) and (b2) $\Gamma=1$, $\ell=1$ and $a=1$; and (c1) and (c2) $\Gamma=1$, $\ell=2$ and $a=-1$. The initial state is set at the center of the system, with a fixed size of $L = 101$. $T_c$ corresponds to the time when $N$ decreases to $10^{-7}$. }
	\label{fig1} 
\end{figure}

The dissipative dynamics of density matrix $\rho(t)$ is described by the Lindblad master equation~\cite{GLindblad,HPBreuer}
\begin{equation}\label{master_eq}
	\frac{\ud\rho(t)}{\ud{}t}
	=\mathcal{L}[\rho(t)]
	=-i\left[H_\tot\rho(t)-\rho(t)H_\tot^\dagger\right]
	+\mathcal{D}[\rho(t)],
\end{equation}
where the Hamiltonian $H_\tot=H_0+V_\nh$ is non-Hermitian, $\mathcal{L}$ is called the Lindbladian superoperator and $\mathcal{D}$ is the dissipation
superoperator
\begin{equation}
	\mathcal{D}[\rho(t)]
	=\Gamma\sum_{m}\left[D_m\rho D_m^\dag
	-\frac{1}{2}\left\{D_m^\dag D_m,\rho\right\}\right].
\end{equation}
which contains a set of jump operators $D_m$ as shown in Eq. (\ref{noise}), all with the same strength $\Gamma$.
Strictly speaking, Eq. (\ref{master_eq}) neglects the terms $c_1 \rho c_1^\dagger$ and $c_L \rho c_L^\dagger$, which affect the occupation of the vacuum state and the correlation between the single-particle state and the vacuum state. However, this does not alter the results in the dynamics of the particle, which is the only aspect we are interested in here (see details in Appendix A).
 We set $\Gamma=1$ and boundary dissipation strength $\gamma=0.5$ without loss of generality.
 Since $\mathcal{L}$ is time-independent, one can express $\rho(t)=e^{\mathcal{L}t}\rho(0)$. In our numerical simulation, we 
adopt the 4th-order Runge-Kutta method to integrate the master equation and thereby calculate the 
time evolution superoperator $e^{\mathcal{L}t}$.

To visually observe the impact of bond dissipation $D_m$ on the relaxation time of the boundary-dissipative system, we first examine the change in particle occupancy over time. A particle is initially placed at the center of the system, i.e., $\rho(0)=|\frac{L+1}{2}\rangle\langle\frac{L+1}{2}|$ (let $L$ be odd). 
For any time $t$, we can record the occupation probability $\rho_{mm}(t)=\braket{m|\rho(t)|m}$ of every site $m$, as shown in Figs.~\ref{fig1}(a1, b1, c1), and calculate the total number of particles $N(t)=\sum_m\rho_{mm}(t)$, as shown in Figs.~\ref{fig1}(a2, b2, c2). The particle will eventually escape the lattice, and all $\rho_{mm}$ as well as $N$ will approach zero as time passes, regardless of whether there is bond dissipation [Figs.~\ref{fig1}(b1, b2, c1, c2)] or not [Figs.~\ref{fig1}(a1, a2)]. However, bond dissipation does accelerate the particle loss process. To quantitatively characterize this acceleration effect, we use a cutoff of total particle number $N=10^{-7}$, that is, we evaluate the system's relaxation time $T_c$ when $N=10^{-7}$. From Figs. ~\ref{fig1}(a2), (b2), and (c2), it can be seen that, compared with the case without bond dissipation, when the dissipation with $\ell=1$ and $a=1$ ($\ell=2$ and $a=-1$) is added, the relaxation time $T_c$ is reduced to approximately $1/20$ ($1/800$) of its original value.

\begin{figure}[t]
	\centering
	\includegraphics[width=0.4\textwidth]{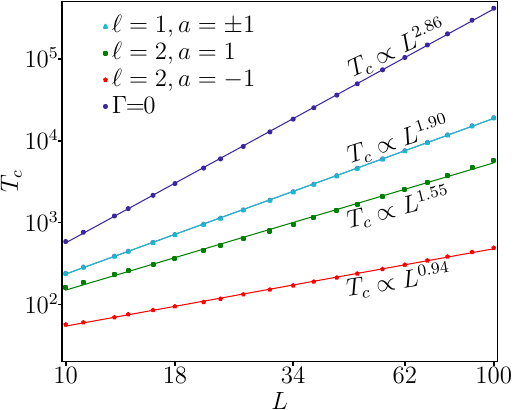}
	\caption{ The relaxation time $T_c$ varies with size $L$. Numerical fitting results: $T_c\sim L^{2.86}$ if $\Gamma=0$; $T_c\sim L^{1.9}$ if $\ell=1$ and $a=\pm1$; $T_c\sim L^{0.94}$ if $\ell=2$ and $a=-1$; and $T_c\sim L^{1.55}$ if $\ell=2$ and $a=1$.}
	\label{fig2}
\end{figure}
  
Next, we study the relationship between the relaxation time $T_c$ and the system size. Fig.~\ref{fig2} shows that regardless of the presence of bond dissipation, the relationship between the relaxation time and the system size for the boundary dissipative system can be expressed as $T_c\sim L^{z}$. When there is no bond dissipation, 
$z$ is approximately $2.86$, which is consistent with the result $z\approx 3$ within the error range~\cite{errorL}. However, when bond dissipation is present, $z$ is significantly less than $2.86$, and the magnitude of $z$ also clearly depends on the specific form of bond dissipation, i.e., on $\ell$ and $a$. We observe that when $\ell=1$, the value of $z$ is approximately the same for $a=1$ and $a=-1$, around 1.9. When $\ell=2$ and $a=-1$,
$z$ is minimized, indicating that in this case, the reduction effect of bond dissipation on the relaxation time is most significant, which is consistent with the results in Fig.~\ref{fig1}.

We now analyze the reason why the particle number conserving bond dissipation $D_m$ can significantly reduce the relaxation time of the boundary-dissipative system. The relaxation time $T_c$ here is inversely proportional to the Liouvillian spectral gap $\Delta_g$, which is defined as the minimum absolute value of the real part of the nonzero eigenvalues of the Liouvillian superoperator. Without bond dissipation, i.e., when $\Gamma=0$, $\Delta_g$ is twice the smallest modulus of the imaginary part of the nonzero eigenvalues of the total Hamiltonian $H_{\tot}$.
We can sort the eigenlevels of $H_{\tot}$ in ascending order of their real parts, which are mainly determined by the eigenlevels of $H_0$ since $V_{\nh}$ can be considered a perturbative term when $L$ is sufficiently large. 
The index of the energy mode is denoted as $n_E$, and then we introduce a size-independent quantity $\epsilon=(n_E-1)/(L-1)$.
Clearly, the smallest, middle and largest real parts of the eigenvalues of $H_{\tot}$ correspond to
$\epsilon=0$, $\epsilon=0.5$ and $\epsilon=1$, respectively.

\begin{figure}[t]
	\centering
	\includegraphics[width=0.485\textwidth]{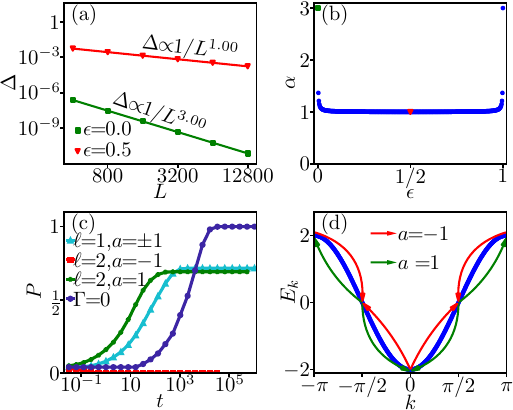}
	\caption{(a) The absolute value of the imaginary part of the eigenvalues with the smallest real part ($\epsilon=0$) and the middle real part ($\epsilon=0.5$) of $H_{tot}$ varies with the system size $L$.
	(b) Inverse scaling coefficient $\alpha$ of $\Delta\propto1/L^\alpha$ for all $\epsilon$.
	(c) Without and with bond dissipation, the change of occupation rate $P$ over time $t$.
	(d) A schematic diagram of the physical effects generated by bond dissipation with 
	$\ell=2$.}
	\label{fig3}
\end{figure}

We examine the absolute value of the imaginary part of each eigenvalue, denoted as $\Delta$, as the system size changes.
Fig.~\ref{fig3}(a) shows $\Delta\propto1/L^{3}$ when the corresponding real part is the smallest and $\Delta\propto1/L$ when the corresponding real part is in the middle of the spectrum. 
It is easy to verify numerically that the relationship between the absolute value of the imaginary part 
$\Delta$ of all eigenvalues and the system size satisfies $\Delta\propto1/L^{\alpha}$. We show the variation of 
$\alpha$ with $\epsilon$ in Fig.~\ref{fig3}(b). It can be seen that $\alpha$ has its highest value of $3$ at the smallest and largest real parts ($\epsilon=0$ and $\epsilon=1$), and its lowest value of $1$ for most $\epsilon$ within 
$(0, 1)$, except for a few levels close to the bottom (top) of the eigenlevels of $H_{\tot}$. This creates a cup shape with a wide, flat bottom, as shown in Fig.~\ref{fig3}(b). It can be conjectured that the previously discovered $T_c\sim L^3$
scaling behavior mainly originates from the states at $\epsilon=0$ and $\epsilon=1$. This is also consistent with Fig.~\ref{fig3}(a), where $\Delta$ at $\epsilon=0$ is significantly smaller than the value at $\epsilon=0.5$, and the relaxation time is determined by the smallest $\Delta$. To further confirm this point, we introduce
\begin{equation}
	P(t)=\left(\braket{\psi_1|\rho(t)|\psi_1}+\braket{\psi_L|\rho(t)|\psi_L}\right)/\Tr\rho(t).
\end{equation}
It describes the ratio of the sum of the particle numbers in the state $\psi_1$ at $\epsilon=0$ and the state 
$\psi_L$ at $\epsilon=1$ to the total particle number at any time $t$~\cite{explain}. As seen in Fig.~\ref{fig3}(c), when there is no bond dissipation (i.e., $\Gamma=0$), after a certain period of time, all states except $\psi_1$ and $\psi_L$ dissipate, and $P$ is approximately equal to $1$. This proves that the relaxation time is determined by the states at $\epsilon=0$ and $\epsilon=1$, hence $T_c\sim L^3$. When bond dissipation with $\ell=2$ and $a=-1$ is added, 
$P$ quickly becomes $0$. Therefore, the states at $\epsilon=0$ and $\epsilon=1$ do not determine its boundary dissipation, which explains its scaling behavior $T_c\sim L^z$ with $z\approx 1$. When the bond dissipation with other parameters is added, 
$P$ eventually stabilizes at a value between $0$ and $1$. Therefore, the relaxation time is influenced by the states at $\epsilon=0$ and $\epsilon=1$, but not solely determined by them. Consequently, the scaling behavior of the relaxation time with size is given by $T_c\sim L^z$, where $z$ lies between $1$ and $3$.

\begin{figure}[!th]
	\centering
	\includegraphics[width=1\linewidth]{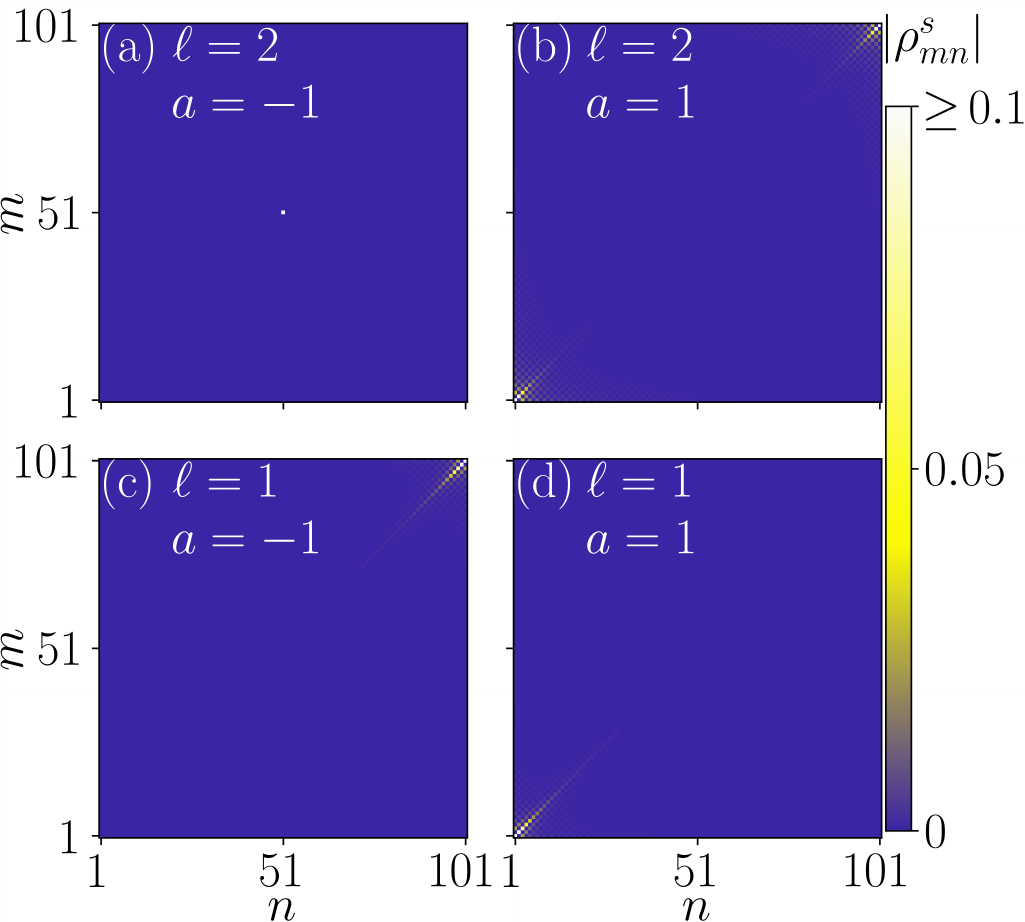}
	\caption{In the eigenbasis of the Hamiltonian $H_0$, the absolute values of the density matrix elements $\rho^s_{mn}$ for steady states with the bond dissipation (a) $\ell=2$ and $a=-1$, (b) $\ell=2$ and $a=1$, (c) $\ell=1$ and $a=-1$, and 
		(d) $\ell=1$ and $a=1$. Here $L=101$, $\gamma=0$, $\Gamma=1$, $m$ and 
		$n$ are the indices of the eigenstates of $H_0$.}
	\label{fig4}
\end{figure}

The behavior of $P$ over time shown in Fig.~\ref{fig3}(c) can be understood through the influence of bond dissipation on the eigenstates of $H_0$. Its eigenvalue is given by $E=-2J\cos(k)$ and wave function is $e^{ikm}$, with $k=2\pi n/L (n\in(-L/2,L/2])$, and the phase difference between the next-nearest neighbor (NNN) lattice sites is $\Delta\phi=2k$. For the states at the bottom ($k=0$) and top ($k=\pi$) of the energy band, the NNN sites are in phase. For the states in the middle of the energy band ($k=\pi/2$), the NNN sites are out of phase. Therefore, when $a=-1$, this bond dissipation annihilates the in-phase states and produces out-of-phase states, as shown by the red line in Fig.~\ref{fig3}(d). This results in only the states near the middle of the spectrum participating in the boundary dissipation behavior, as indicated by $P=0$ in Fig.~\ref{fig3}(c).  When $a=1$, this bond dissipation annihilates the out-of-phase states and produces in-phase states, as shown by the green line in Fig.~\ref{fig3}(d). Therefore, $P$ is a non-zero value, but it is also never equal to $1$. To clearly illustrate this, we set the boundary dissipation strength $\gamma=0$ and examine the regulatory effect of bond dissipation in the eigenbasis of $H_0$. 
We can calculate the eigenstates and their corresponding eigenvalues of the Liouvillian superoperator 
$\mathcal{L}$. The steady state $\rho^{s}$, defined as $\rho^{s}=\lim_{t\rightarrow\infty}\rho(t)$, corresponds to the zero eigenvalue, i.e., $\mathcal{L}[\rho^{s}] = 0$.
We express $\rho^{s}$ in the eigenbasis of $H_0$, as shown in Fig.~\ref{fig4}. The manipulation effects of the bond dissipation with different parameters $\ell$ and $a$ are obvious here. From Fig.~\ref{fig4}(a), we observe that the bond dissipation with $\ell=2$ and $a=-1$ indeed drives the steady state to primarily occupy the middle of the energy spectrum, consistent with the results shown in Fig.~\ref{fig3}(c, d). When $\ell=2$ and $a=1$, the steady state predominantly occupies the states at the edges of the energy spectrum, which is also consistent with the discussion in Fig.~\ref{fig3}(c, d). We note that the steady state occupies many states, not just the highest and lowest ones, and therefore, for the case of $\ell=2$ and $a=1$, $P$ in Fig.~\ref{fig3}(c) is not equal to $1$.  Similarly, for $\ell=1$ and $a=\pm 1$, although bond dissipation drives the steady state to occupy the edges of the energy spectrum, it still includes many states, as shown in Fig.~\ref{fig4}(c,d). Thus, the sum of the proportions of the lowest and highest states is not equal to $1$ [Fig.~\ref{fig3}(c)].
Since the relaxation time of boundary dissipation is much longer than the time for the system to reach a steady state due to bond dissipation, it can be assumed that bond dissipation has already distributed the states before the boundary dissipation process begins. The distribution of states is such that the proportion of each state is the same as the steady-state distribution caused by bond dissipation when $\gamma=0$. Although boundary dissipation reduces the total number of particles, this proportion remains constant. Overall, the particles in the states with $\alpha=1$ are more easily dissipated, but this bond dissipation forces the proportions of each state to remain constant. This leads to particles from the $\alpha=3$ states transitioning into the $\alpha=1$ states to maintain the constant proportions, as shown in Fig.~\ref{fig3}(c). This, in turn, results in the shortening of the relaxation time. 

\section{In the presence of AL}
\begin{figure}[t]
	\centering
	\includegraphics[width=0.485\textwidth]{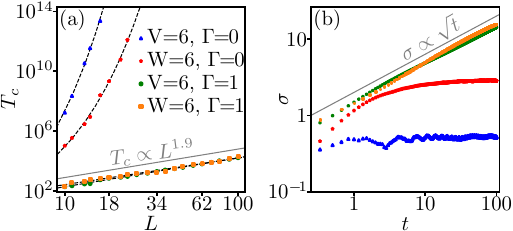}
	\caption{ (a) The change of relaxation time with system size both with and without bond dissipation. When $\Gamma=0$, to smooth the data, we averaged over $1000$ samples for both disordered and quasiperiodic systems. For the quasiperiodic system, each sample corresponds to an initial phase $\theta$. When $\Gamma\neq 0$, taking multiple samples has little effect on the results. When $\Gamma=1$, the relationship between $T_c$ and $L$ takes a power-law form, with $T_c \sim L^z$, where $z \approx 1.78$ when $W=6$ and $z \approx 1.93$ when $V=6$. (b) The evolution of $\sigma$ over time. $\kappa\approx 0.48$ for $V=6,\Gamma=1$ and $\kappa\approx 0.52$ for $W=6,\Gamma=1$. Here we take $\beta=(\sqrt{5}-1)/2$ and the bond dissipation with $\ell=1$ and $a=1$. The result is similar when $\beta$, $\ell$ and $a$ take other values. 
	}
	\label{fig5}
\end{figure}

Finally, we consider the impact of bond dissipation on the relaxation time of boundary-dissipative systems in the presence of AL. AL can be induced by either a random on-site potential 
\begin{equation}
	V_\mathrm{r} = \sum_{m=1}^LV_m c_m^\dagger c_m,
\end{equation}
where $V_m$ is uniformly distributed in $[-W/2,W/2]$ with $W$ being the disorder strength, or a
quasiperiodic potential
\begin{equation}
	V_\mathrm{qp} = V\sum_{m=1}^L\cos(2\pi\beta{}m+\theta)c_m^\dagger c_m,
\end{equation}
where $\beta$ is an irrational number, and $V$ and $\theta$ are the strength and initial phase of the quasiperiodic potential, respectively. $H=H_0+V_\mathrm{qp}$ is the Aubry-Andr\'{e} model~\cite{AA}, which exhibits an Anderson transition at $V/J=2$. When $V/J>2$ ($V/J<2$), all eigenstates are localized (extended). When the added potential is random disorder, i.e., 
$V_\mathrm{qp}$ is replaced with $V_\mathrm{r}$, any weak disorder strength $W$ can cause the system to become localized.
In the absence of bond dissipation, the relaxation time undergoes an exponential scaling relation with the change in system size $T_c\propto{}e^{\eta{}L}$ in the AL phase~\cite{Prosen,Zhou2022exp}, as shown in Fig.~\ref{fig4}(a).
This is because an electron localized within the bulk of the lattice has an exponentially small chance, relative to $L$, of reaching the particle loss channel at the lattice boundaries.
When bond dissipation $D_m$ is introduced, we observe that the scaling behavior of the relaxation time changes to a power-law form as the system size varies [Fig.~\ref{fig5}(a)]. This cannot be explained by the previous property of $D_m$ selectively targeting specific states, because now all states are localized in the absence of $D_m$. This indicates that this bond dissipation may disrupt the localized nature of the states. To verify this, we set the strength of boundary dissipation 
$\gamma=0$ and study the impact of bond dissipation on the wave packet dynamics of this localized system. A common quantity used to describe the dynamics of wave packet evolution is the mean square displacement~\cite{Hiramoto,Geisel1997,Roati,WangYc2020}
\begin{equation}
	\sigma(t) = \sqrt{\sum_m[m-(L+1)/2]^2\rho_{mm}(t)},
\end{equation}
which measures the width of the wave packet initially located at the center of the system, with $L$ taken as an odd number.
After a period of time, the change of $\sigma$ over time can be expressed as $\sigma\sim t^{\kappa}$, where the dynamical index $\kappa$ corresponds to different types of diffusion: $\kappa=0$, $\kappa<1/2$, $\kappa\approx 1/2$, $\kappa>1/2$, and $\kappa=1$ for localized, subdiffusive, normal diffusive, superdiffusive, and ballistic diffusion, respectively.
From Fig.~\ref{fig5}(b), when there is no bond dissipation, the system is localized, but with the addition of bond dissipation, the wave packet evolution approaches normal diffusion.
This explains why the scaling relation of the relaxation time with system size changes from an exponential form to a power-law form. Additionally, for the quasi-periodic system and the random disorder system, $\kappa$ is approximately $0.48$ and $0.52$, respectively. This means that diffusion in the disordered system is slightly faster, leading to a shorter time to reach a steady state. As a result, the value of $z$ in the relationship $T_c \sim L^z$ for the quasi-periodic system ($z \approx 1.93$ in Fig.~\ref{fig5}(a)) is slightly larger than that for the disordered system ($z \approx 1.87$ in Fig.~\ref{fig5}(c)).

\section{Conclusion and Discussion}
We have investigated the impact of bond dissipation shown in Eq. (\ref{noise}) on the relaxation time of boundary-dissipative systems, and found that bond dissipation can significantly reduce the relaxation time. The scaling of the relaxation time $T_c\sim L^{z}$ changes from $z=3$ to a value significantly less than $3$. This is because bond dissipation can selectively target specific states, thereby eliminating or reducing the influence of the states with the longest relaxation times. For Anderson localized systems, bond dissipation can change the scaling behavior of the relaxation time from an exponential form to a power-law form as the system size varies. This is because bond dissipation disrupts the localization properties of the system, making it more akin to normal diffusive behavior. Our results highlight the significant role of bond dissipation in manipulating the relaxation processes of quantum systems.

For simplicity, we mainly discussed the case where both boundaries are particle loss channels. If one side is gain and the other is loss, the results we present regarding the impact of bond dissipation on the relaxation time still apply. Additionally, we primarily discussed the case where the boundary dissipation strength $\gamma<1$. In the Appendix, we present the results for $\gamma=1$ and $\gamma>1$. We can observe that states at the center of the energy spectrum seem to exhibit a phase transition-like behavior when $\gamma=1$. This change can be measured through the bond dissipation with $\ell=2$ and $a=-1$. Our results also introduce some interesting questions worthy of further investigation. For instance, how does bond dissipation affect the relaxation time of boundary-dissipative systems in the presence of interactions? How does this impact change in the presence of many-body localization? Can this bond dissipation also regulate the relaxation time in systems with other types of dissipation?

\begin{acknowledgments}
This work is supported by National Key R\&D Program of China under Grant No.2022YFA1405800, the National Natural Science
Foundation of China (Grant No.12104205), the Key-Area Research and Development Program of Guangdong Province (Grant No. 2018B030326001), Guangdong Provincial Key Laboratory (Grant No.2019B121203002).
\end{acknowledgments}

\appendix
\section{Discussions on Eq. (4)}

The evolution of a quantum state in a Markovian reservoir is described by the Lindblad equation
\begin{equation}
	\frac{d\rho}{dt}=-i[H,\rho]+\sum_{\mu\in(bulk,boundary)}(2L_{\mu}\rho L_{\mu}^{\dagger}-\{L_{\mu}^{\dagger}L_{\mu},\rho\}).
\end{equation}
In this system, there are two types of reservoirs: the bulk (bond) dissipation, as in Eq. (3) in the main text, and the boundary dissipation:
\begin{equation}\label{losseq}
	L_{1}=\sqrt{\gamma}c_1,~~~~~~~~~~L_{2}=\sqrt{\gamma}c_L.
\end{equation}
Here, the bulk dissipation does not break particle number conservation, while the boundary dissipation describes particle loss. In the long-time limit, all particles are absorbed into the boundary reservoir.

In this work, we focus on the single-particle relaxation dynamics. The Hilbert space is composed of single-particle states  $|j\rangle=c_{j}^{\dagger}|0\rangle$ with $j=1,2,\cdots,L$ and the vacuum state $|0\rangle$. By incorporating the terms $-c_{1}^{\dagger}c_{1}\rho-c_{L}^{\dagger}c_{L}\rho-\rho c_{1}^{\dagger}c_{1}-\rho c_{L}^{\dagger}c_{L}$ into the effective Hamiltonian $H_{\tot}=H_0-i\gamma c_{1}^{\dagger}c_{1}-i\gamma c_{L}^{\dagger}c_{L}$, the Lindblad equation becomes
\begin{widetext}
\begin{equation}\label{MEeq}
	\frac{d\rho}{dt}=-i(H_{\tot}\rho-\rho H_{\tot}^{\dagger})+\sum_{\mu\in bulk}(2L_{\mu}\rho L_{\mu}^{\dagger}-\{L_{\mu}^{\dagger}L_{\mu},\rho\})+\tilde{\mathcal{D}}(\rho),
\end{equation}
\end{widetext}
where $\tilde{\mathcal{D}}(\rho)=2\gamma c_{1}\rho c_{1}^{\dagger}+2\gamma c_{L}\rho c_{L}^{\dagger}$. In our work, we neglect the term $\tilde{\mathcal{D}}(\rho)$. The single particle Hilbert space $\{|j\rangle\}$ is not fixed with the vacuum state $|0\rangle$, and the total particle number is the trace of the density matrix, $N=tr(\rho)$. The term $\tilde{\mathcal{D}}(\rho)$ will bring the single-particle states $|1\rangle$ and $|L\rangle$ to the vacuum state, generating the occupation of vacuum state and the correlation between the single particle state and the vacuum state. Since we are only interested in the dynamics of the particle, the occupation of the vacuum state and the correlation between the single particle and vacuum states are not important. Therefore, we assume that the boundary particle loss is described by a non-Hermitian Hamiltonian. Of course, directly using Eq. (\ref{losseq}) and Eq. (\ref{MEeq}) for the calculations would also yield the results in our paper, but some discussions would become less convenient. In our main text, when the strength of bond dissipation $\Gamma = 0$, the Liouvillian spectral gap is twice the smallest absolute value of the imaginary part of the nonzero eigenvalues of the total Hamiltonian $H_{\tot}$. Therefore, in the discussions related to Fig. 3, we primarily base our analysis on $H_{\tot}$. Because the non-Hermitian term $V_{\text{nh}}$ in $H_{\tot}$ only acts on the two edge sites, it can be considered a perturbation, and the real part of $H_{\tot}$ is determined by $H_{0}$. Based on this, we sort the real parts of the eigenvalues of $H_{\tot}$ in ascending order and then analyze the relationship between the absolute values of the imaginary parts of each eigenvalue and the system size. This allows us to understand the influence of different states on the relaxation time in the spectrum and, in turn, the dissipation process, as well as why bond dissipation accelerates this process. If we were to consider $\tilde{\mathcal{D}}$, this analysis would become much more complicated. Therefore, given that $\tilde{\mathcal{D}}$ does not alter our results, we did not include this term in the main text.

\begin{figure}[t]
	\centering
	\includegraphics[width=0.485\textwidth]{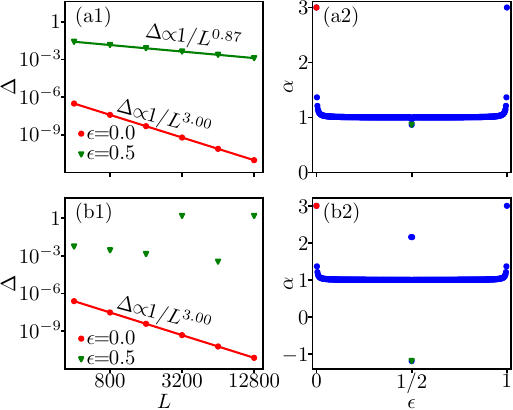}
	\caption{Scaling of the absolute $\Delta$ of imaginary part of eigenlevels at $\epsilon=0$ 
		and $\epsilon=0.5$ of  the total Hamitonian $H_{\tot}$ for (a1) $\gamma=1$ and (b1) $\gamma=2$;
		and inverse scaling coefficient $\alpha$ of $\Delta\propto1/L^\alpha$ versus $\epsilon$
		for  (a2) $\gamma=1$ and (b2) $\gamma=2$.
	}
	\label{figs1}
\end{figure}

\section{Boundary dissipation strength $\gamma\geq 1$}

In the main text, we have choose the particle-loss strength $\gamma=0.5$ to
avoid the case where $\gamma\ge1$. When $\gamma=1$, the dependence of the absolute value of the imaginary part at 
$\epsilon=0.5$ on the system size follows $\Delta\sim 1/L^{\alpha}$, with $\alpha$ being less than $1$, approximately $0.87$ [see Fig.~\ref{figs1}(a1)]. By comparing Fig.~\ref{figs1}(a2) with Fig. 3(b) in the main text, it can be seen that only $\alpha$ at $\epsilon=0.5$ decreases, while the rest remains the same as for $\gamma<1$.
When $\gamma>1$, the dependence of the absolute value of the imaginary part at $\epsilon=0.5$ on the system size no longer follows $\Delta\sim 1/L^{\alpha}$, but instead exhibits large oscillations [see Fig.~\ref{figs1}(b1)]. These oscillations occur only at $\epsilon=0.5$ [see Fig.~\ref{figs1}(b2)]. We emphasize that when $\gamma\geq1$, the change in the absolute value of the imaginary part at $\epsilon=0.5$ is difficult to detect without bond dissipation because the relaxation time is mainly determined by the eigenvalues near $\epsilon=0$ and $\epsilon=1$. Therefore, the change caused by 
$\gamma$ does not affect the scaling relationship $T_c\sim L^3$. However, when bond dissipation with 
$\ell=2$ and $a=-1$ is introduced, the system's relaxation time is primarily determined by the eigenvalues near 
$\epsilon=0.5$. At this point, the changes brought by $\gamma\geq1$ can be detected.



\begin{thebibliography}{99}
\bibitem{explain} We provide a simple understanding that $\alpha$ equals $3$ at the spectral boundary and $1$ at the spectral center. In the free fermionic model Eq. (\ref{freeH}) under open boundary conditions, the eigenvalues are $E_{j}=-2J\cos\frac{j\pi}{L+1}$, and the corresponding eigenvectors are $|\psi_{j}\rangle=\sqrt{\frac{2}{L+1}} \sum_{m=1}^{L}\sin\frac{jm\pi}{L+1}|m\rangle$ with $j=1,2,\cdots,L$ being the index of eigenstate. The boundary density of states are $D_j=|\psi_{j,1}|^2+|\psi_{j,L}|^2=\frac{4}{L+1}\sin^2\frac{j\pi}{L+1}$. Then from the perturbative perspective, we can show $\Delta_j=-2\gamma D_j$ is proportional to the boundary density of states. Hence near the band edge $j\rightarrow 1$ or $L$, $\sin^2\frac{j\pi}{L+1}\sim L^{-2}$, so $\Delta\sim L^{-3}$, while inside the band $j\rightarrow\frac{L}{2}$, we have $\Delta\sim L^{-1}$.
\end{thebibliography}
\end{document}